\documentclass[conference]{IEEEtran}

\usepackage{ifpdf}
\ifpdf

\ifCLASSOPTIONcompsoc
  \usepackage[nocompress]{cite}
\else
  \usepackage{cite}
\fi
\ifCLASSINFOpdf \else \fi

\usepackage{amsmath}
\usepackage{algorithmic}
\usepackage{array}
\usepackage{bm}
\usepackage[pdftex]{graphicx}
\usepackage{subfigure}
\usepackage{multirow}
\usepackage{booktabs}
\usepackage{indentfirst}
\usepackage{amsfonts,amssymb}
\usepackage{lineno}
\usepackage{color}
\usepackage{algorithm}
\usepackage{algorithmic}
\usepackage{makecell}

\renewcommand\thesection{\Roman{section}}

\begin{document}

\title{Online Learning Based NLOS Ranging Error Mitigation in 5G Positioning}

\author{
	\centerline{Jiankun Zhang and Hao Wang}\\
    \IEEEauthorblockA{
        Huawei Technologies Co., Ltd, Beijing 100095, China\\
		Email: \{zhangjiankun4, hunter.wanghao\}@huawei.com}
}

\maketitle

\begin{abstract}
The fifth-generation (5G) wireless communication is useful for positioning due to its large bandwidth and low cost. However, the presence of obstacles that block the line-of-sight (LOS) path between devices would affect localization accuracy severely. In this paper, we propose an online learning approach to mitigate ranging error directly in non-line-of-sight (NLOS) channels. The distribution of NLOS ranging error is learned from received raw signals, where a network with neural processes regressor (NPR) is utilized to learn the environment and range-related information precisely. The network can be implemented for online learning free from retraining the network, which is computationally efficient. Simulation results show that the proposed approach outperforms conventional techniques in terms of NLOS ranging error mitigation.
\end{abstract}

\IEEEpeerreviewmaketitle

\section{Introduction}

The fifth-generation (5G) wireless communication presents unique opportunities for positioning on user equipments (UE), due to its large bandwidths, arrays with many antenna elements and favorable propagation conditions [1,2]. With 5G positioning, UE can support integrated sensing and communication via a joint signal processing framework, making full use of spectrum resources. Applications of 5G positioning include smart factory, personal navigation, as well as autonomous vehicles [3]. However, practical deployment of 5G positioning systems still faces a number of critical challenges, such as non-line-of-sight (NLOS) propagation [4]. NLOS propagation results in positively biased range estimates, which in turn degrades localization performance. NLOS conditions occur frequently in many practical harsh environments, including inside buildings, in urban canyons or under tree canopies. Therefore, it is imperative to understand the impact of NLOS conditions on localization systems, and to develop techniques that mitigate their effects.

Machine learning methods are recently introduced to address the NLOS mitigation problem for their ability to accumulate knowledge from data, which can be coarsely classified as two categories: The first class is NLOS ranging error regression, the goal of which is to reduce the effect of the ranging error in NLOS conditions. In [5], ranging error due to NLOS effects was estimated through support vector machines and Gaussian processes regressors (GPR). [6] proposed a ranging method based on kernel principal component analysis, in which the selected channel parameters are projected onto a nonlinear orthogonal high-dimensional space for ranging. A probabilistic learning approach was introduced in [7] to mitigate the ranging error due to NLOS propagation by combining variational inference with probabilistic neural networks.

In many cases, it may not be possible to detect an NLOS condition with full certainty. In that case, the second class of NLOS mitigation - soft NLOS mitigation can be taken, that is, the probability of NLOS condition is estimated. In this class, NLOS identification and error mitigation are then combined into one single step, and the ranging likelihood function becomes a mixture of LOS and NLOS models. In [8], soft range information (SRI) was introduced to represent probable range values, which are derived through machine learning techniques. In [9], a Bayesian probabilistic approach for NLOS localization was developed using multivariate Gaussian mixture models (GMM). [10] proposed a deep learning approach with two neural modules to generate SRI of ranging error directly from received NLOS raw signals. In [11], the simultaneous range error mitigation and environment identification were presented. In [12], NLOS components were demonstrated to provide position and orientation information that consequently increase position estimation accuracy in 5G millimeter wave systems. However, most of the aforementioned approaches assume a fixed, often pre-trained model for the relationship between the inter-node measurement and ranging error. Though enjoying a significant improvement in performance using full waveforms, such deep learning methods suffer from generalization problems and are prone to overfitting. For UE-based positioning task, surrounding environment changes rapidly. When UE moves to a new environment, the neural network needs to be re-trained which results in additional computation cost and time delay.

In this paper, we propose a novel online learning approach for soft ranging error mitigation in NLOS channel, which benefits 5G positioning. In many cases, it may not be possible to determine an NLOS bias with full certainty. Instead, the distribution of NLOS ranging error (represented by a mean and a variance) is given in the proposed paradigm. In particular, the proposed approach is implemented by a network with neural processes regressor (NPR). According to the variational inference theory, the aggregator and decoder of the network can learn the environment and range-related information respectively. Therefore, the NLOS ranging error can be directly learned from the measured power delay profile (PDP). An online learning procedure is designed free from retraining the network, which is computationally efficient and suitable for implementation on a user equipment (UE). Simulation results show that the proposed approach outperforms conventional techniques in terms of NLOS ranging error mitigation.

This paper is organized as follows. Section II presents the channel model and problem formulation. Section III shows the theoretical analysis for the NPR algorithm. Section IV introduces the network architecture and learning procedure. In Section V, numerical simulations are carried out to investigate the performance of the proposed scheme. Finally, concluding remarks are given in Section VI.

\section{System Model}

Our goal is to estimate the range using the received positioning reference signal (PRS) in a 5G downlink system. First, we have measured the complex channel impulse response (CIR) of the channel $h(t) = {\sum_{k}{a_{k}\delta\left( {t - \tau_{k}} \right)}}$, where $a_k$ and $\tau_k$ are the amplitude and delay of the $k$-th path respectively. Then we obtain its squared amplitude $|h(t)|^2$, known as power delay profile (PDP). The PDP can also be represented by a vector containing the power and delay of all the paths, i.e., $\textbf{r} = [|a_0|^2, \cdots, |a_{N-1}|^2, \tau_0, \cdots, \tau_{N-1}]^T$, where $N$ is the number of paths. If we suppose the distance between the base station (BS) and user equipment (UE) is $d$, the estimated distance can be represented as
\begin{equation}
\displaystyle{\overset{\_}{d} = c\tau_{0} = \left\{ \begin{matrix}
{d + n_{\text{LOS}},} & {\text{if}~H = \text{LOS}} \\
{d + g\left( \mathbf{r} \right) + n_{\text{NLOS}},} & {\text{if}~H = \text{NLOS}} \\
\end{matrix} \right.,}
\end{equation}
where $c$ is the light speed, $H \in \left\{ \text{LOS},~\text{NLOS} \right\}$ is the propagation condition, $\left. n_{\text{LOS}} \right.\sim\mathcal{N}\left( 0,\sigma_{\text{LOS}}^{2} \right)$ and $\left. n_{\text{NLOS}} \right.\sim\mathcal{N}\left( 0,\sigma_{\text{NLOS}}^{2} \right)$ are the additive noise in LOS and NLOS conditions, respectively. Here, $g\left( \mathbf{r} \right)$ can be regarded as a random process that NLOS ranging error satisfies when PDP spectrum is known. For a single measurement, $g\left( \mathbf{r} \right)$ is a random variable subject to a specific distribution. It means that the distance to be measured $d$ can be considered as a random variable with unknown distribution. According to the Bayesian criterion, the posterior probability of $d$ is given by
\begin{equation}
\begin{aligned}
p\left( d \middle| \mathbf{r} \right) & = {\sum\limits_{H \in {\{{\text{LOS},\text{NLOS}}\}}}\frac{p\left( {\mathbf{r},H} \right)p\left( d \middle| {\mathbf{r},H} \right)}{p\left( \mathbf{r} \right)}} \\
& = {\sum\limits_{H \in \{\text{LOS},\text{NLOS}\}}{p\left( H \middle| \mathbf{r} \right)p\left( d \middle| {\mathbf{r},H} \right)}}
\end{aligned}
\end{equation}
which can be further written as
\begin{equation}
\begin{aligned}
p\left( d \middle| \mathbf{r} \right) & = p\left( {H = \text{LOS}} \middle| \mathbf{r} \right)p\left( d \middle| {\mathbf{r},H = \text{LOS}} \right) \\
& + p\left( {H = \text{NLOS}} \middle| \mathbf{r} \right)p\left( d \middle| {\mathbf{r},H = \text{NLOS}} \right).
\end{aligned}
\end{equation}

Let us take the LOS condition as an example. The conditional distribution can be represented as
\begin{equation}
\displaystyle{p\left( d \middle| {\mathbf{r},H = \text{LOS}} \right) = \frac{p\left( \mathbf{r} \middle| {d,H = \text{LOS}} \right)p(d)}{\int\nolimits_{0}^{\infty}{p\left( \mathbf{r} \middle| {d,H = \text{LOS}} \right)p(d)\text{d}d}}.}
\end{equation}
Note that the denominator of (4) is the normalization factor that enables $\int_{0}^{\infty}{p\left( d \middle| {\mathbf{r},H = \text{LOS}} \right)\text{d}d = 1}$. The conditional distribution can be further given by
\begin{equation}
\displaystyle{p\left( d \middle| {\mathbf{r},H = \text{LOS}} \right) = \left( \frac{1}{2\pi\sigma_{\text{LOS}}^{2}} \right)^{\frac{1}{2}}e^{- \frac{{({d - \overset{\_}{d}})}^{2}}{2\sigma_{\text{LOS}}^{2}}}.}
\end{equation}
Therefore we can derive
\begin{equation}
\displaystyle{\mathbb{E}\left\{ d \middle| \mathbf{r},H = \text{LOS} \right\} = \overset{\_}{d}.}
\end{equation}
Similarly, for the NLOS condition, we have
\begin{equation}
\displaystyle{\mathbb{E}\left\{ d \middle| \mathbf{r},H = \text{NLOS} \right\} = \overset{\_}{d} - E\left\{ {g\left( \mathbf{r} \right)} \right\}.}
\end{equation}

By modeling the distance as a random variable, the estimation of distance can be derived by means of the minimum mean square error (MMSE) estimator as
\begin{equation}
\begin{aligned}
{\hat{d}}_{\text{MMSE}} & = \mathbb{E}\left\{ d \middle| \mathbf{r} \right\} \\
& = p\left( {H = \text{LOS}} \middle| \mathbf{r} \right) \mathbb{E}\left\{ d \middle| \mathbf{r},H = \text{LOS} \right\} \\
& ~~~ + p\left( {H = \text{NLOS}} \middle| \mathbf{r} \right)\mathbb{E}\left\{ d \middle| \mathbf{r},H = \text{NLOS} \right\} \\
& = \overset{\_}{d} - ~p\left( {H = \text{NLOS}} \middle| \mathbf{r} \right) \cdot \mathbb{E}\left\{ {g\left( \mathbf{r} \right)} \right\}
\end{aligned}
\end{equation}

According to (8), we can design the NLOS ranging error mitigation scheme with two-stage structure, as shown in Fig. 1. First, the probability of NLOS condition can be derived according to the received PDP, which is $p\left( {H = \text{NLOS}} \middle| \mathbf{r} \right)$. Second, the mean value of NLOS ranging error can be obtained through the received PDP, which is $\mathbb{E}\left\{ {g\left( \mathbf{r} \right)} \right\}$. In the following part of this paper, we concentrate on the second part, i.e., precise estimation of ranging bias $g\left( \mathbf{r} \right)$.

\begin{figure} [t]
\centering
\includegraphics[width=0.5\textwidth]{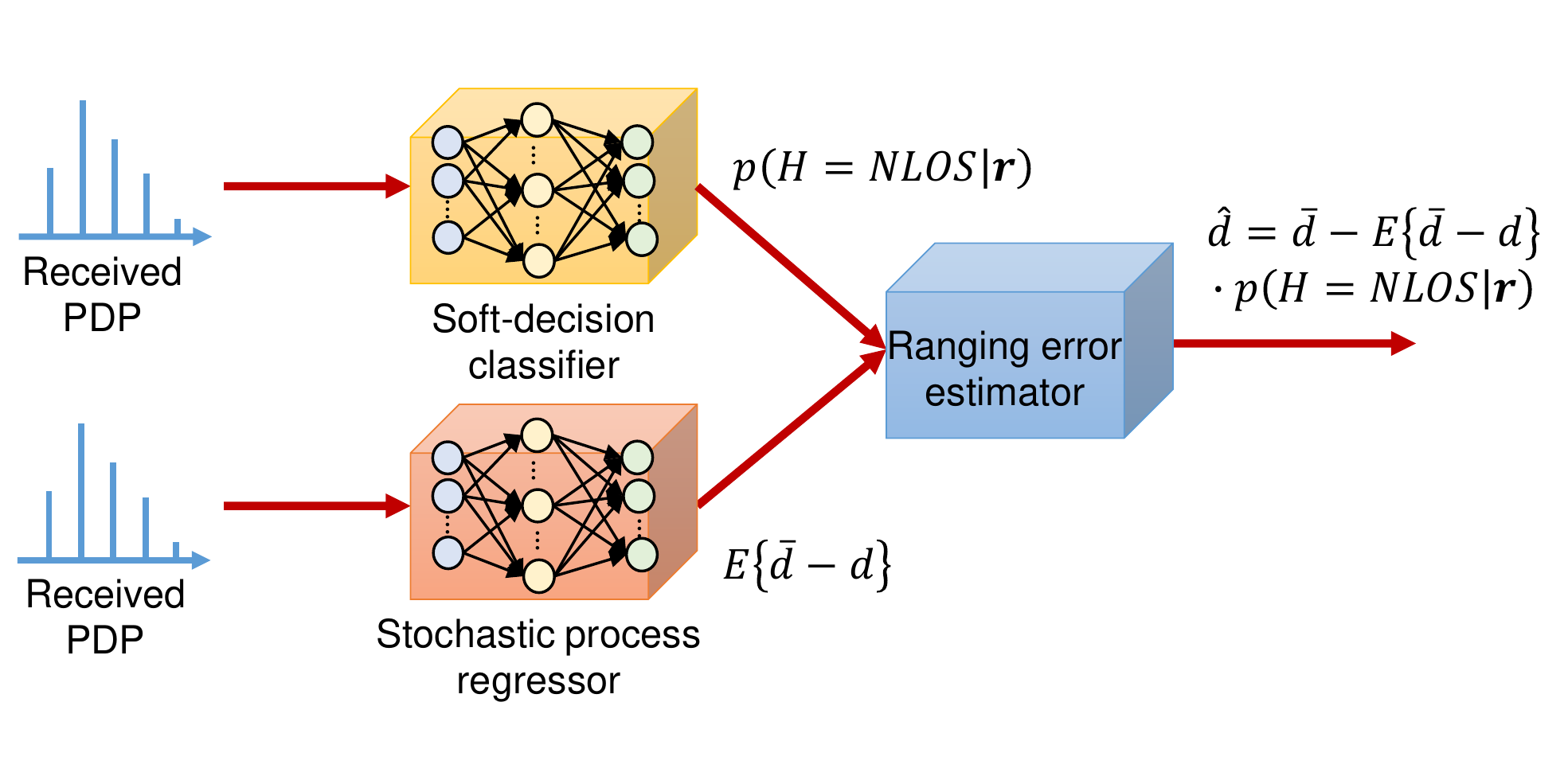}
\caption{System diagram of two-stage NLOS ranging error estimator.}
\label{fig}
\end{figure}

\section{Theoretical Principle of NPR}

As a fundamental description of the channel, the received waveform represents how a signal propagates from the transmitter to the receiver in a multipath channel. As hinted before, received waveforms inherently contain both range-related and environment-related information. In this section, we develop a Bayesian probabilistic algorithm, called neural processes regression (NPR), which has the following advantages:

\begin{itemize}
\item {NPR is reasonable and interpretable, where the aggregator and decoder can learn the environment-related and range-related information, respectively.}
\end{itemize}
\begin{itemize}
\item {NPR is capable of rapid adaption to new environment, which is computationally efficient during online learning without re-training the network.}
\end{itemize}
\begin{itemize}
\item {NPR can learn the posterior distribution of the desired output, and can estimate the uncertainty in the predictions.}
\end{itemize}

NPR was first proposed in [13] by DeepMind, which combines the advantages of data-efficient Gaussian process (GP) and computation-efficient neural networks (NN). Given $n$ context points and $m$ target points, inference with trained neural processes correspond to a forward pass in a deep NN, which scales with $O(n+m)$ as opposed to the $O((n+m)^3)$ runtime of classic GPs. Furthermore the model overcomes many functional design restrictions by learning an implicit kernel from the data directly [14,15]. In this work, we extend NPR to an online learning approach, which is suitable for precise positioning of UE.

Again, we can model the estimation of ranging error as the following tasks
\begin{equation}
\displaystyle{\Delta d = \overset{\_}{d} - d = g\left( \mathbf{r} \right) + n_{F},}
\end{equation}
where $\left. n_{F} \right.\sim\mathcal{N}\left( 0,\sigma_{F}^{2} \right)$ is the quantization noise whose variance is known a priori and determined by the system bandwidth. Since $\Delta d$ and $\mathbf{r}$ are not strictly one-to-one correspondence, we assume that $g\left( \mathbf{r} \right)$ follows Gaussian process, namely $\left. g\left( \mathbf{r} \right) \right.\sim\mathcal{G}\mathcal{P}\left( \mathbf{m},\mathbf{K} \right)$.

Suppose that we have a training data set $\mathcal{T}_{N} = \left\{ {{\Delta d}_{n},\mathbf{r}_{n}} \right\}_{n = 1}^{N}$, as well as a single measurement result $\mathbf{r}_{\text{test}}$. By defining $\mathbf{g}_{N} = \left( {g\left( \mathbf{r}_{1} \right),\cdots,g\left( \mathbf{r}_{N} \right)} \right)$, $\mathbf{R}_{N} = \left( {\mathbf{r}_{1},\cdots,\mathbf{r}_{N}} \right)$, and ${\mathbf{\Delta}\mathbf{d}}_{N} = \left( {{\Delta d}_{1},\cdots,{\Delta d}_{N}} \right)$, then the likelihood function of the training set is given by
\begin{equation}
\begin{aligned}
p\left( {\mathbf{\Delta}\mathbf{d}}_{N} \middle| \mathbf{g}_{N} \right) & = {\prod\limits_{n = 1}^{N}{p\left( {\Delta d}_{n} \middle| {g\left( \mathbf{r}_{n} \right)} \right)}} \\
& = {\prod\limits_{n = 1}^{N}{\mathcal{N}\left( {\Delta d}_{n};g\left( \mathbf{r}_{n} \right),\sigma_{F}^{2} \right)}},
\end{aligned}
\end{equation}
and the posterior distribution of ${\mathbf{\Delta}\mathbf{d}}_{N}$ can be represented as
\begin{equation}
\displaystyle{p\left( {\mathbf{\Delta}\mathbf{d}}_{N} \middle| \mathbf{R}_{N} \right) = {\int{p\left( \mathbf{g}_{N} \middle| \mathbf{R}_{N} \right)p\left( {\mathbf{\Delta}\mathbf{d}}_{N} \middle| \mathbf{g}_{N} \right)d\mathbf{g}_{N}}}.}
\end{equation}
By substituting (10) into (11), we can get
\begin{equation}
\displaystyle{p\left( {\mathbf{\Delta}\mathbf{d}}_{N} \middle| \mathbf{R}_{N} \right) = {\int{p\left( \mathbf{g}_{N} \middle| \mathbf{R}_{N} \right){\prod\limits_{n = 1}^{N}{\mathcal{N}\left( {\Delta d}_{n};g\left( \mathbf{r}_{n} \right),\sigma_{F}^{2} \right)}}d\mathbf{g}_{N}}}.}
\end{equation}

The key concept of NPR is that the random process $g$ can be parameterised by a random variable $z$, and written as $g\left( \mathbf{r}_{n} \right) = f\left( {\mathbf{r}_{n},z} \right)$ for fixed and learnable function $f$. It also means that the randomness in $g$ is due to that of $z$. The generative model of (12) then follows
\begin{equation}
\displaystyle{p\left( {\mathbf{\Delta}\mathbf{d}}_{N} \middle| \mathbf{R}_{N} \right) = {\int{p(z){\prod\limits_{n = 1}^{N}{\mathcal{N}\left( {\Delta d}_{n};f\left( {\mathbf{r}_{n},z} \right),\sigma_{F}^{2} \right)}}dz}}.}
\end{equation}
Since the function $f\left( {\mathbf{r}_{n},z} \right)$ is non-linear, we can use variational inference to learn it. Let $q\left( z \middle| {\mathbf{\Delta}\mathbf{d}_{\mathbf{N}},\mathbf{R}_{\mathbf{N}}} \right)$ be a variational posterior of the latent variables $z$, parameterised
by another neural network that is invariant to permutations of the sequences $\mathbf{R}_{\mathbf{N}}$. Then the evidence lower-bound (ELBO) is given by
\begin{equation}
\begin{aligned}
{\ln{p\left( \mathbf{\Delta}\mathbf{d}_{\mathbf{N}} \middle| \mathbf{R}_{\mathbf{N}} \right)}} & = {\ln{\sum\limits_{z}{p\left( z,\mathbf{\Delta}\mathbf{d}_{\mathbf{N}} \middle| \mathbf{R}_{\mathbf{N}} \right)}}} \\
& = {\ln{\sum\limits_{z}{\frac{p\left( z,\mathbf{\Delta}\mathbf{d}_{\mathbf{N}} \middle| \mathbf{R}_{\mathbf{N}} \right)}{q\left( z \middle| {\mathbf{\Delta}\mathbf{d}_{\mathbf{N}},\mathbf{R}_{\mathbf{N}}} \right)}q\left( z \middle| {\mathbf{\Delta}\mathbf{d}_{\mathbf{N}},\mathbf{R}_{\mathbf{N}}} \right)}}} \\
& \geq \mathbb{E}_{q{({z|{\mathbf{\Delta}\mathbf{d}_{\mathbf{N}},\mathbf{R}_{\mathbf{N}}}})}}\left\lbrack {\ln\frac{p\left( z,\mathbf{\Delta}\mathbf{d}_{\mathbf{N}} \middle| \mathbf{R}_{\mathbf{N}} \right)}{q\left( z \middle| {\mathbf{\Delta}\mathbf{d}_{\mathbf{N}},\mathbf{R}_{\mathbf{N}}} \right)}} \right\rbrack.
\end{aligned}
\end{equation}
Finally, we can derive
\begin{equation}
\begin{aligned}
& {\ln{p\left( \mathbf{\Delta}\mathbf{d}_{\mathbf{N}} \middle| \mathbf{R}_{\mathbf{N}} \right)}} \geq \\
& \mathbb{E}_{q{({z|{\mathbf{\Delta}\mathbf{d}_{\mathbf{N}},\mathbf{R}_{\mathbf{N}}}})}}\left\lbrack {\ln{p\left( \mathbf{\Delta}\mathbf{d}_{\mathbf{N}} \middle| z,\mathbf{R}_{\mathbf{N}} \right) + {\ln\frac{p(z)}{q\left( z \middle| {\mathbf{\Delta}\mathbf{d}_{\mathbf{N}},\mathbf{X}_{\mathbf{N}}} \right)}}}} \right\rbrack,
\end{aligned}
\end{equation}
where the first term of the right side is the optimized objective function, and the second term is Kullback-Leibler (KL) divergence.

The physical meaning of (15) is clear: our ultimate goal is to maximize the likelihood of observations $p\left( \mathbf{\Delta}\mathbf{d}_{\mathbf{N}} \middle| \mathbf{R}_{\mathbf{N}} \right)$ to solve the corresponding parameters, but the likelihood can only be obtained by integrating the implicit variables, i.e., $\sum\limits_{z}{p\left( \mathbf{\Delta}\mathbf{d}_{\mathbf{N}} \middle| \mathbf{R}_{\mathbf{N}},z \right)}$, which is usually not computable. For example, if the hidden variable is discrete, there is a combinatorial explosion problem; while if the hidden variable is continuous, there is a problem that its distribution is unknown or the integral is not obtainable. Then the variational distribution $q\left( z \middle| {\mathbf{\Delta}\mathbf{d}_{\mathbf{N}},\mathbf{R}_{\mathbf{N}}} \right)$ is introduced to approximate a priori $p(z)$, and $\ln{p\left( \mathbf{\Delta}\mathbf{d}_{\mathbf{N}} \middle| \mathbf{R}_{\mathbf{N}} \right)}$ can be obtained according to the above derivation, then maximizing ELBO becomes an approximate optimization goal.

\begin{figure} [t]
\centering
\includegraphics[width=0.5\textwidth]{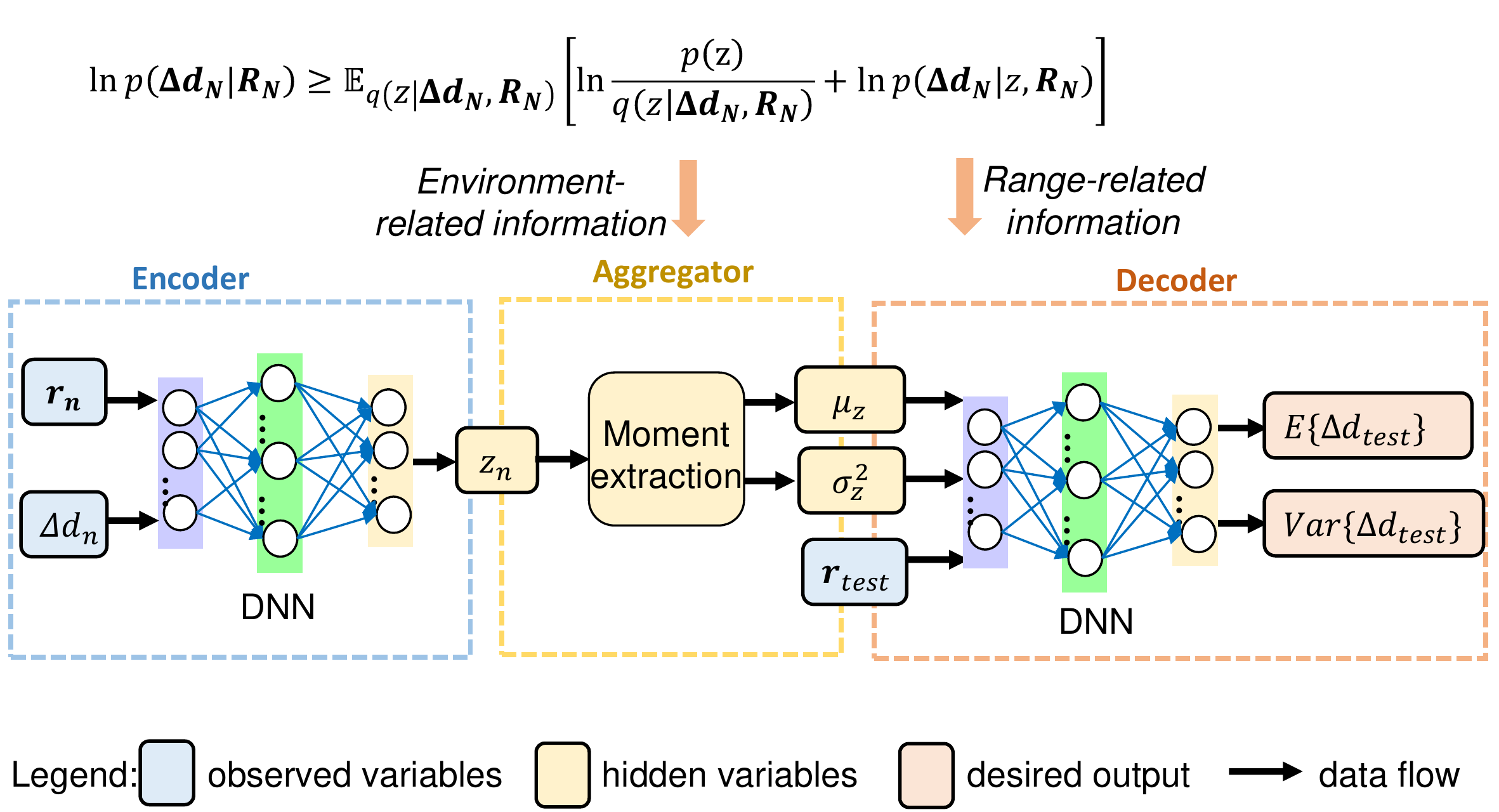}
\caption{Network structure of stochastic processes regressor using NPR.}
\label{fig}
\end{figure}

\section{Network Implementation of Online Learning}

\subsection{Network Architecture}

According to the above theory, an NPR network with the structure shown in Fig. 2 can be designed. The network includes three parts: encoder, aggregator and decoder. The specific structure and functions of each module are as follows:

\begin{itemize}
\item {\textbf{Encoder}: Mapping the data from the input space to the presentation space. The inputs are $\left( {\mathbf{x} = \mathbf{R}_{N},\mathbf{y} = {\mathbf{\Delta}\mathbf{d}}_{N}} \right)_{n}$. The output is $z_{n} = h\left( \left( {\mathbf{x},\mathbf{y}} \right)_{n} \right)$, where $h$ is realized by neural network fitting.}
\end{itemize}
\begin{itemize}
\item {\textbf{Aggregator}: Getting the statistics of $z$, and treat $\left. z \right.\sim\mathcal{N}\left( {\mu_z,\sigma_z^{2}} \right)$ as Gaussian distribution. In this module, the input is $z_ n$. The output is its mean value $\mu_z = \frac{1}{N}{\sum\nolimits_{n = 1}^{N}z_{n}}$ and variance $\sigma_z^2 = \frac{1}{N}{\sum\nolimits_{n = 1}^{N}(z_n-\mu_z)^2}$. The use of aggregator reduces the network complexity from $O(N^3)$ to $O(N)$.}
\end{itemize}
\begin{itemize}
\item {\textbf{Decoder}: Obtaining the output of the random process. The inputs of this module is $\mathbf{r}_\text{test}$, $\mu_z$ and $\sigma_z^{2}$, and the outputs are $\mathbb{E}\left\lbrack f\left( {\mathbf{r}_\text{test},z} \right)\right\rbrack$ and $\mathbb{V}\text{ar}\left\lbrack f\left( {\mathbf{r}_\text{test},z} \right)\right\rbrack$. $f$ is also realized by neural network fitting.}
\end{itemize}

One advantage of designing the network in this way is that each part of the network just conforms to the variational inference model. In particular, the encoder learns how to extract the environment-related information from the received signals; the aggregator extracts the environment-related information $p(z)$; while decoder extracts the range-related information $p\left( \mathbf{\Delta}\mathbf{d}_{\mathbf{N}} \middle| z,\mathbf{R}_{\mathbf{N}} \right)$ from the test data and hidden variables. As will be described in detail in Section IV.B, when the UE moves to a new environment, the aggregator will extract the environment-related information by updating the statistics of $z$, while the encoder and decoder are without the need to be trained.

\subsection{Learning Procedure}

As mentioned before, an advantage of the NPR scheme is that one can learn the environment-related information by collecting the statistics of hidden variable $z$ through aggregator. Since the networks in encoder and decoder only learn range-related information, online learning can be carried out without the need to re-train the networks. In order to achieve these objectives, the NPR networks are trained and used as follows, as shown in Fig. 3.

\begin{itemize}
\item {\textbf{Encoder Training}: In the off-line pre-training phase, the NPR is treated as an autoprecoder and trained end-to-end in a supervised manner. More specifically, a dataset of the measured PDPs in different NLOS environments are collected and the encoder and decoder are trained to be able to predict the ranging bias given the input PDP. It is important to mention here that through the end-to-end training of the auto-precoder model, the encoder in Fig. 3(a) learns in an unsupervised way how to optimize its compressive hidden variable $z$.}
\end{itemize}
\begin{itemize}
\item {\textbf{Decoder Training}: Since the training label in the above process is the real value of $\Delta d$, while the output of the decoder is the distribution of $\Delta d$; therefore an additional decoder training process is needed here, as shown in Fig. 3(b) and 3(c). In the decoder pre-training phase, we fix the parameters of the networks, generate random seeds according to $\left. z \right.\sim\mathcal{N}\left( {\mu_z,\sigma_z^{2}} \right)$, and obtain the mean and variance of $\Delta d_\text{test}$. Then $(\mu_z,\sigma_z^{2},\mathbb{E}\{\Delta d_\text{test}\},\mathbb{V}\text{ar}\{\Delta d_\text{test}\})$ are used as labels to train the parameters of the decoder until the network converges.}
\end{itemize}
\begin{itemize}
\item {\textbf{Online Testing}: After the NPR network is trained, it is decoupled into two parts in the online-testing (prediction) phase: (i) The encoder and aggregator learn the environment-related information and update $(\mu_z,\sigma_z^{2})$ in the aggregator. Here, the input of the encoder comes from the LOS condition. In particular, if the NLOS identify module (see Fig. 1) considers the scene as a LOS scene ($P(H = \text{NLOS} | \mathbf{r})$ less than a threshold), the group of data will be used as an online label. Here the input $\mathbf{r}$ is the PDP after removing the first (LOS) path, and $\Delta d = \Delta\tau c$, where $\Delta\tau$ is the time of arrival (TOA) difference between the first path and the second path, and $c$ is the light speed. (ii) The PDP measurements, as well as the compressive statistics of $z$, will be inputted to the decoder network and used to directly predict the distribution of ranging error.}
\end{itemize}

\begin{figure} [t]
\centering
\includegraphics[width=0.5\textwidth]{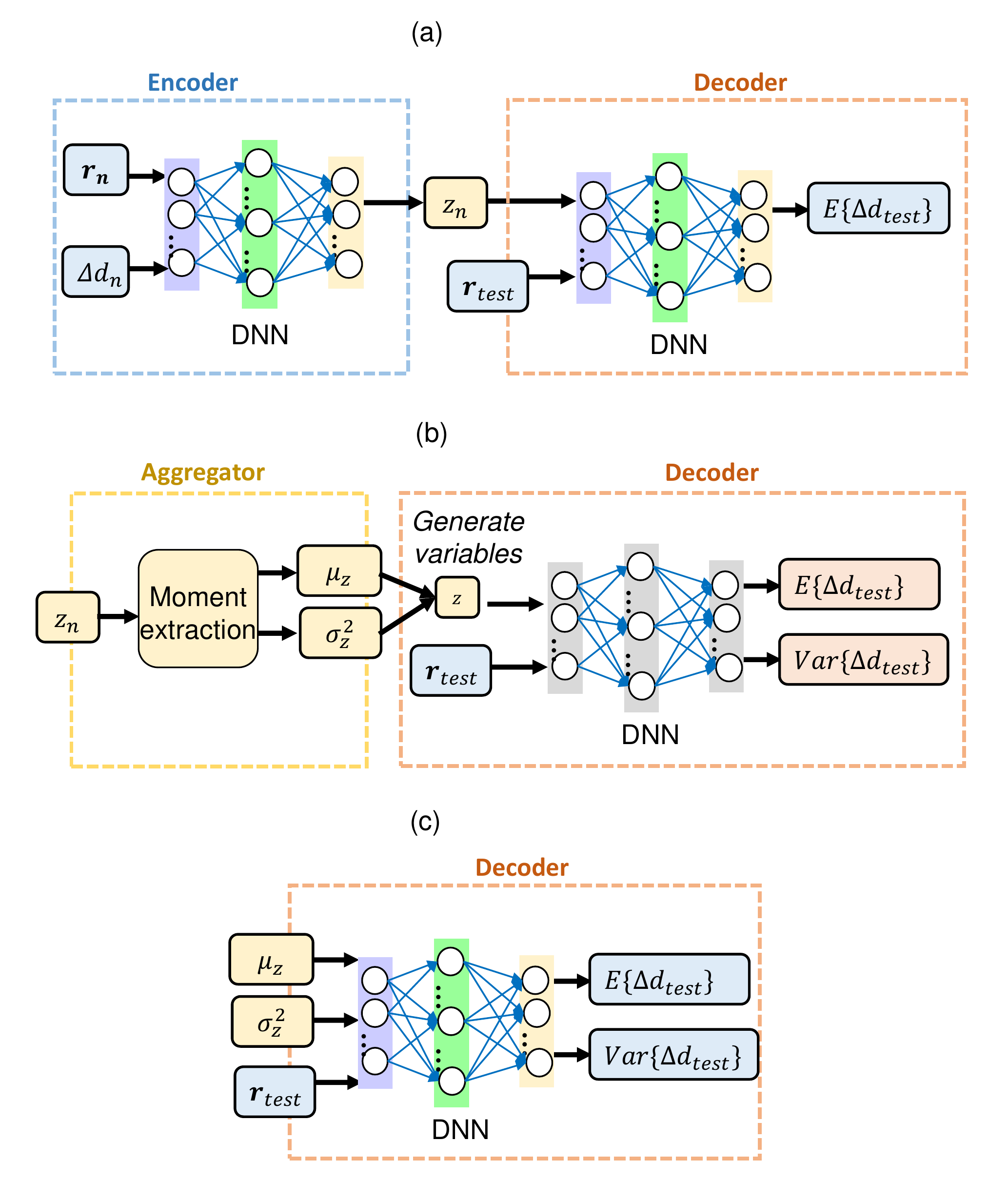}
\caption{Learning procedure of online NPR: (a) encoder training; (b) generate random seeds; (c) decoder training. The grey background of the DNN implies the network is fixed in the current step.}
\label{fig}
\end{figure}

\section{Simulation}

\subsection{Implementation Details}

\begin{table}[t]
\centering
\caption{Key Parameters in Simulations}
\label{table}
\begin{tabular}{c||c}
\hline
\textbf{Parameter} & \textbf{Value} \\
\hline
Carrier frequency & 28 GHz \\
Bandwidth & 400 MHz \\
Subcarrier spacing & 120 kHz \\
Reference signal & Positioning reference signal (PRS) \\
PRS period & 10 ms \\
Num. of BS antennas & $12\times16$ \\
Num. of UE antennas & $1\times4$ \\
Total subframes	& 10000 \\
\hline
\end{tabular}
\end{table}

\begin{figure} [t]
\centering
\includegraphics[width=0.4\textwidth]{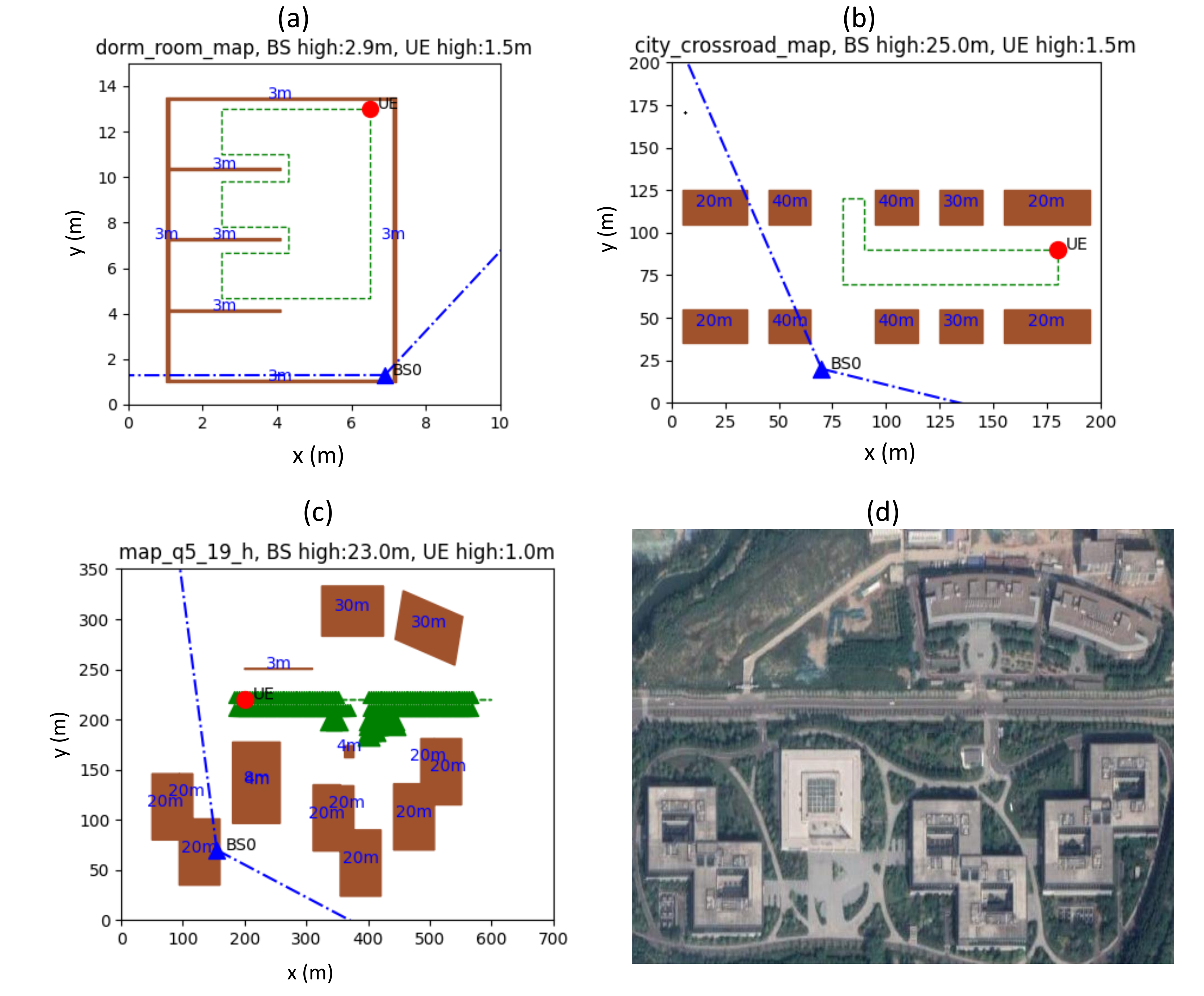}
\caption{(a)-(c) Examples of different maps in 5G datasets generated from ray-tracing model,
and (d) real channel corresponding to map(c) where data are collected.}
\label{fig}
\end{figure}

In this section, we provide numerical examples for ranging estimation for NLOS channel in a 5G simulation link. The parameters of our simulation are summarized in Table I. In particular, the BS is equipped with $12\times16$ antennas, and the UE is assumed to employ $4\times1$ antennas. We consider a single-RF-chain receiver with a carrier frequency of 28 GHz (corresponding to frequency range 2), a total bandwidth of 400 MHz with 120 kHz subcarrier spacing. The PRS period is set to 10 ms according to 3GPP standard. A cyclic prefix of length 7 is used. 64 pilot OFDM symbols are sent, for a total duration of 3.52 ms. The whole datasets consist of 10 different maps generated using ray-tracing model, where each map corresponds to an individual dataset, as shown in Fig. 4. More specifically, the environment of the link we apply has been calibrated from the real measurements of indoor and urban channel, where UE moves and blocked by different obstacles.

For all datasets, We utilize 80\% of the data samples for training and the rest 20\% for testing, as a commonly used strategy for data assignment in deep learning methods. In our simulation, the NPR network is implemented in Pytorch. The computing platform is two 14-core 3.00GHz Intel Xeon E5-2690 CPUs (28 Logical Processors) with GPU NVIDIA Quadro P2000. The training data consists of a number of groups randomly chosen from static scenarios in the ray-tracing model. We train the network with 10,000 epochs. At each epoch, the training and validation sets contain 5,000 and 1,000 samples, respectively. For test sets, the neural network of the channel encoder is directly implemented in the analog circuits. The NPR is trained using the stochastic gradient descent method and Adam optimizer. The learning rate is set to be 0.001. The batch size is set to 1000. In our experiment settings, we choose the $L_2$ loss as the cost function.

The compared algorithms include two typical NLOS bias estimators, GPR [5] and down sampling with residual block (we abbreviate it as DS-RB) [10]. Specifically, the DS-RB estimator consists of a initial linear layer, 3 down-sampling blocks, a residual block and a linear output layer, which is the same as in [10]. The hyper-parameters used in GPR and DS-RB are the same as those mentioned in the original papers, respectively. The ranging error without any NLOS mitigation techniques is also shown as a benchmark. The cumulative distribution function (CDF) of the ranging error is used to measure the performance of these estimators.

\begin{figure} [t]
\centering
\includegraphics[width=0.4\textwidth]{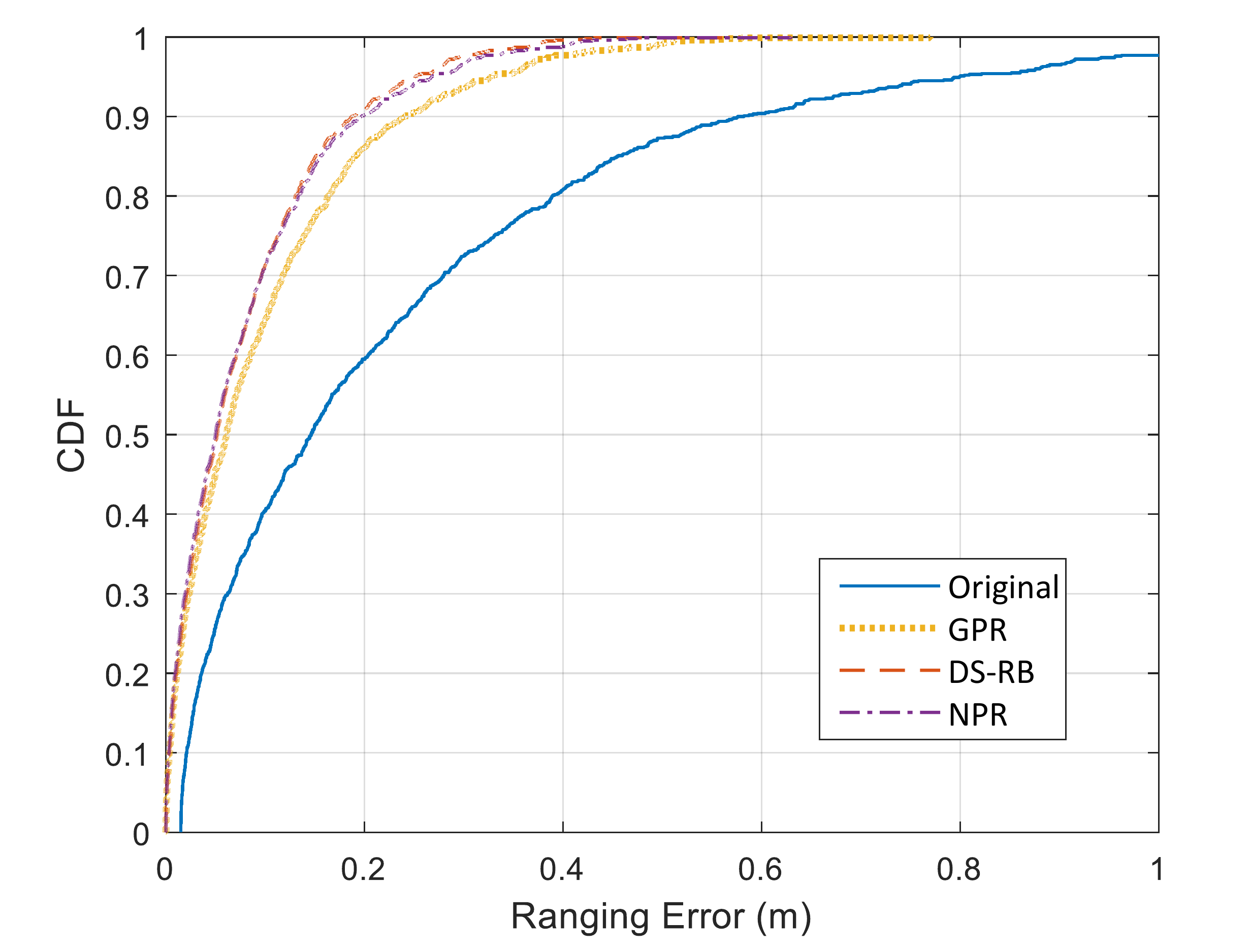}
\caption{The CDFs of the ranging error on testing data trained on the same datasets.}
\label{fig}
\end{figure}

\begin{figure} [t]
\centering
\includegraphics[width=0.4\textwidth]{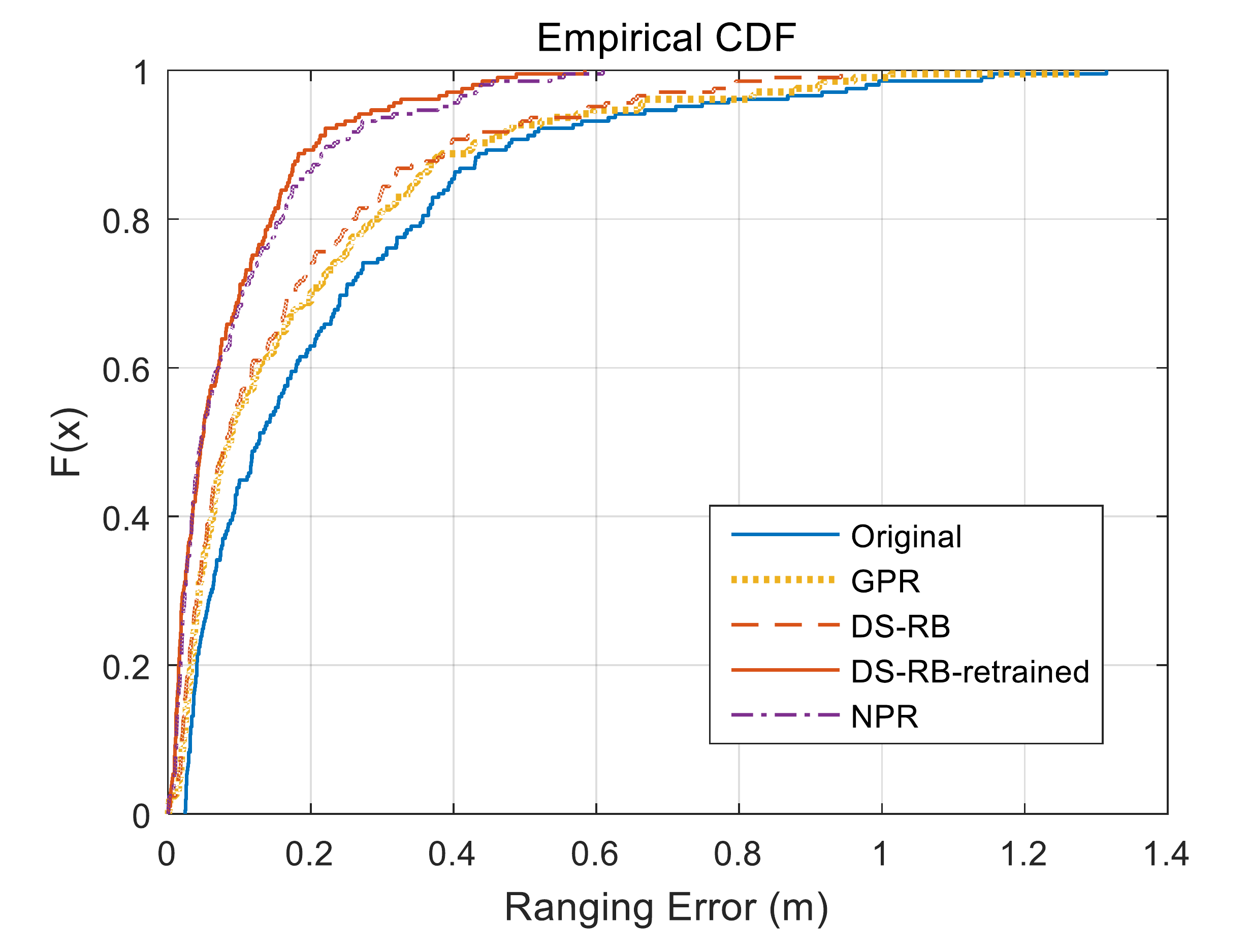}
\caption{The CDFs of the ranging error on testing data trained on different datasets.}
\label{fig}
\end{figure}

\subsection{Simulation Results}

First, we train and test all the estimators in five different datasets, and show the performance of different algorithms for ranging error regression. Figure 5 shows the CDF of the ranging error at the off-line learning stage. It can be observed that NPR and DS-RB have comparable performance, and both of them outperform GPR. Such a performance gain comes from the benefits of the two-stage structure, as well as a better learning of implicit metrics directly from data. One of the benefit of NPR is its functional flexibility as they can learn `kernels' from data directly. Once the model is trained it can regress more than just one dataset, and it will produce sensible results for curves generated using any kernel parameters observed during training.

\begin{table}[t]
\centering
\caption{Quantitative Results on Online Range Error Mitigation}
\label{table}
\begin{tabular}{c|c|c|c|c}
\hline
Technique & $P_{0.1}$ [m] & $P_{0.5}$ [m] & $P_{0.9}$ [m] & $T_{\text{on}}$ [ms] \\
\hline
Original & 0.031 & 0.124 & 0.478 & - \\
\hline
GPR [5] & 0.023 & 0.096 & 0.458 & 8.52 \\
\hline
DS-RB [10] & 0.021 & 0.083 & 0.449 & 2.63 \\
\hline
DS-RB-retrained & 0.011 & 0.054 & 0.229 & 177 \\
\hline
NPR & 0.013 & 0.054 & 0.260 & 3.12 \\
\hline
\end{tabular}
\end{table}

Finally, we evaluate the online learning performance of range error mitigation for NPR. The CDFs of different algorithms are shown in Fig. 6. In this case, all the estimators are trained under five different datasets as mentioned above, and are tested under a new dataset different from all the trained datasets. For DS-RB, results of retraining the network in each new environment are also given out and shown as DS-RB-retrained. Quantitative results on all the datasets are presented in Table II, where $P_{0.1}$, $P_{0.5}$ and $P_{0.9}$ indicate the ranging bias for CDF = 10\%, 50\%, and 90\%, respectively. It can be observed that NPR-based technique achieves higher accuracy than conventional techniques without network retraining. In particular, stochastic process regression based on NPR can realize a centimeter-level accuracy, which can reduce the ranging error at 50\% CDF of about 26.7\% with respect to the baseline without NLOS mitigation, and of about 52.4\% and 78.2\% with respect to DS-RB and GPR, respectively. Since the estimators are not pre-trained in the test scenario, The performance of DS-RB decreases due to mismatch between data and model, resulting in large estimated values and fails to conduct effective mitigation task. By comparison, the proposed approach achieves better results in all datasets, implying both effectivess and generality. The outstanding performance of NPR comes from its well-designed structure where the aggregator can learn the environment-related information without retraining the network.

Table II also shows the average online processing time (denoted as $T_{\text{on}}$) of different learning algorithms. It can be observed that NPR has a comparable performance with DS-RB in terms of online running time, which is much lower than GPR and DS-RB-retrained. Although retrained DS-RB can achieve better accuracy than NPR, the complexity is much higher. In practice, training deep-learning models online is a computationally intensive task, often requiring hours or even days for large models. In this case the computational complexity of online-training would be unacceptable. Moreover, with the increasing of data samples $N$ for online training, The computational complexity of GPR follows $O(N^3)$ and increases rapidly, while the computational complexity of NPR is restricted to $O(N)$. Therefore, according to our experiments, NPR is more appropriate positioning method for varying environments, such as implemented on the UE.

\section{Conclusion}

In this paper, an online learning approach for ranging error mitigation in NLOS channel is proposed for precise 5G positioning. unlike previous works that determine an NLOS bias with full certainty, the distribution of NLOS ranging error is given out in the proposed paradigm. In particular, the proposed approach is implemented by a network with NPR. According to the variational inference theory, the aggregator and decoder of the network can learn the environment and range-related information respectively. Therefore, the NLOS ranging error can be directly learned from the measured PDP. An online learning procedure is designed free from retraining the network, which is computationally efficient and suitable for implementation on the UE. Simulation results show that the proposed approach outperforms conventional techniques in terms of NLOS ranging error mitigation.

\end{document}